# Raveling the Role of Dopants on Charge Carrier Kinetics of TiO$_2$ Electrodes using Electrochemical Impedance Spectroscopy

Manish K Vishwakarma[1,*], Monojit Bag[2], Puneet Jain[1]

[1]Department of Physics, Indian Institute of Technology Roorkee, Roorkee, Uttarakhand, India 247667

[2]Advanced Research in Electrochemical Impedance Spectroscopy Laboratory, Indian Institute of Technology Roorkee, Uttarakhand, India 247667

[*]Address correspondence to E-mail**:** *mvishwakarma@ph.iitr.ac.in*

**Abstract:**

We synthesized the pure and co-doped titanium dioxide (TiO$_2$) electrodes via spin coating. We examined the optical and electronic properties of as-prepared thin film electrodes with co-doping of transition metals and non-metals. The co-doping of Cu, Zn, and N increase the absorption of the radiation in the visible region. The doping leads to the formation of defect states in the electrodes. In this article, we have studied the carrier kinetics in pristine and co-doped TiO$_2$ electrodes. To study the role of dopants in carrier transport of the synthesized TiO$_2$ based electrodes, the electrochemical impedance spectroscopy measurement is performed in the frequency range of $10^{-1}$ Hz to $10^6$ Hz. The study reveals the influence of dopants on electron-hole recombination in the defect sites present in bulk and the transport mechanism of the electrons and ions to the surface/interface of the electrodes.

## Introduction

TiO$_2$ is the supreme candidate among the wide bandgap metal oxide semiconductors for many applications due to its prodigious properties [1, 2]. It is suitable for photocatalytic and water splitting [3] applications because of its impressive oxidizing capability, chemical stability, non-toxicity, and biological inactiveness. TiO$_2$ is also stable against photo-chemical corrosion. TiO$_2$ have significant applications in photovoltaics and light-emitting diodes (LEDs) as transparent anodic electrodes and gas sensors [3-7]. TiO$_2$ is chosen for anodic electrode material applications due to its low mass density and structural stability against many charging and discharging cycles [8]. The critical issue with pristine TiO$_2$ thin film is low photo efficiency due to its wide bandgap structure. The change in optical properties via doping is encouraged to improve the photo-absorption performance. The doping and co-doping of metallic (Cu, Fe, Ni, Co, Zn, Mn) and non-metallic elements (F, N, C, S, B) are the fundamental way to modify the bandgap as well as improve electronic properties by introducing defects and intermediate states [1, 9-11].

Improved photocatalytic efficiency is linked with the total number of charge carriers reaching the surface of the electrodes, so we must ensure that many charge carriers reach the surface before recombination. Thus, there is a strong need to study the electron and ions transport in the electrode material and how the dopants affect their kinetics. Conduction of ions towards the electrode/electrolyte interface and dielectric behavior of material plays an important role in the low-frequency range, so it is crucial to study low-frequency carrier kinetics in detail. Electrochemical Impedance Spectroscopy (EIS) is a magnificent technique for studying material's charge carrier kinetics and electrical properties [12]. EIS is smart enough to distinguish between the capacitive and resistive response of the working electrode interfaces based on the frequency-dependent current response over some selected range of voltage [13, 14].

In this article, we discuss the change in the optical properties of TiO$_2$ electrodes implemented via co-doping of Cu, Zn, and N. We observed the increment in the visible region photon absorption with the help of the UV-Visible absorption spectrum. The report studies charge carrier kinetics in the TiO$_2$ based electrodes utilizing Electro-impedance spectroscopy. The change in capacitive and resistive behavior of the electrodes is observed with different doping elements (Table-S1). We can study the behavior of the electrodes with the help of simple RC circuit combinations. The resistive behavior of pure and co-doped TiO$_2$ working electrodes for energy-based applications is linked with the transfer of electrons, and capacitive behavior reveals the faradaic and non-faradaic current flow through the interface. The resistance is attributed to electron transfer reactions like electron and hole recombination in the defect centers or catalysis on the active surface sites. The behavior of the resistance and capacitance are different over the broad frequency range ($10^6$ Hz to $10^{-1}$ Hz). We discussed the behavior of the electrodes with the doping using the Nyquist plots and circuit fitting.

## Experimental section

### Chemicals

The chemicals were purchased from Sigma-Aldrich Chemical Co. and used in good and stable condition as received. Fluorine-doped Tin Oxide (FTO) coated glass was used as a thin film coating substrate. Substrates were cut into $1\ cm \times 1\ cm$ size, cleaned thoroughly with DI water, acetone, and isopropyl alcohol (IPA) and dried in hot air. The conductivity of the FTO coated glass surface was $\leq 20$ Ωm, supplied by Shilpent Nagpur, India.

### Electrode synthesis

Pristine and co-doped $TiO_2$ thin films were prepared using the spin coating technique. Precursors were prepared for the p-$TiO_2$ electrodes by dissolving 1 ml of titanium (IV) tetra-isopropoxide (TTIP, $Ti\{OCH(CH_3)_2\}_4$) in 5 ml of isopropyl alcohol (IPA) and stirred on a magnetic stirrer for 5 min. Then 0.5 ml of acetic acid ($CH_3COOH$) was mixed with the solution dropwise to prevent the precipitation of the $TiO_2$ and stirred vigorously for 1 hour. Copper sulphate pentahydrate ($CuSO_4 \cdot 5H_2O$), Zinc Acetate ($(CH_3COO)_2Zn$), and Urea ($NH_2CONH_2$) were used for the Cu, Zn, and N doping, respectively. By dissolving 4 wt.% $CuSO_4 \cdot 5H_2O$, 4wt.% $(CH_3COO)_2Zn$ in methanol separately and 4 wt.% Urea in IPA. The prepared dopant solution is mixed accordingly with the precursor and coated on the FTO coated glass substrate (1 cm × 1 cm) with the help of the spin coater at 3000 rpm for 60 sec. Coated films were dried on the hot plate at $100°C$ for 0.5H and then annealed in a muffle furnace at $450°C$ in the air for 1H to achieve the crystallinity and remove the organic solvents (Fig. 1).

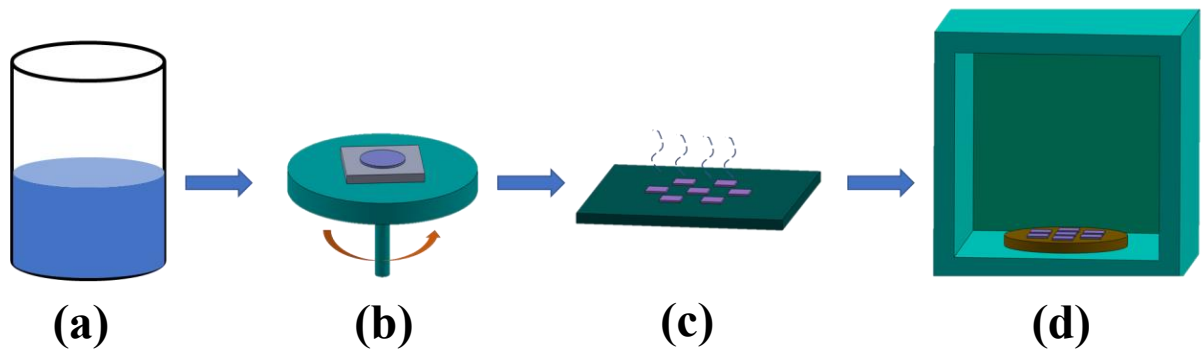

**Fig. 1** Thin Film Fabrication Process **(a)** Precursor, **(b)** Spin coating, **(c)** Drying of samples over hot Plate at $100\ °C$ for 0.5H, **(d)** Annealing of samples in muffle furnace at $450\ °C$ for 1H.

## Thin film characterization

XRD patterns of the pristine and co-doped TiO₂ electrodes were measured with the Smart Lab, Rigaku's thin film x-ray diffractometer using HyPix-3000 high energy resolution 2D HPAD detector. The instrument uses Cu target ($K_\alpha$; $\lambda = 1.54$ Å) and is operated at 45 kV, 200 mA. The data was collected from $20° < 2\theta < 70°$ angle with the scan rate 2°/min. UV-Visible absorbance spectra were taken of all the samples with an Agilent Cary 100 UV-Vis spectrophotometer. The spectra were collected over the 200-800 nm wavelength, referenced against air background. A field emission scanning electron microscope (FE-SEM; Carl Zeiss Ultra, Carl Zeiss; Germany) equipped with energy dispersed X-ray spectroscopy (EDS) was used for the study of thin film's surface morphology and elemental color mapping. EIS measurements of the samples were performed using AutoLab Electrochemical Workstation under dark conditions at 0.1 V bias. The electrolyte (NaCl) of 0.5 M concentration was prepared by dissolving the NaCl in the DI water. In the EIS measurements, an AC signal of 20 mV amplitude was varied in the frequency range from $10^6$ Hz to $10^{-1}$ Hz. The measured impedance data were fitted using ZSimpWin software. Characterization of all the samples were performed in the ambient air conditions.

## Result and discussion

### X-ray diffraction

To check the crystalline quality of as-fabricated electrodes, we have measured X-ray diffraction patterns for pristine TiO₂, Cu, Zn-TiO₂ and Cu, Zn, TiO₂-N electrodes. XRD pattern of the pristine and co-doped $TiO_2$ exclusively manifest the anatase (tetragonal) phase after the annealing (Fig. 2(a)). Two dominant peaks of TiO₂ anatase were observed at $2\theta = 25.3°, 48.1°$ corresponding to $(101), (200)$ miller planes, respectively, also reported by other authors. We observed that intensity of the peaks significantly reduced in co-doped TiO₂ comparative to pristine TiO₂ with the variation in FWHM. The crystallite size of doped and undoped samples were calculated with the help of Scherrer's equation,

$$D = \frac{0.9\lambda}{\beta Cos\theta} \quad \ldots (1)$$

Where D is the crystallite size, λ is the wavelength of the X-Ray, β is the FWHM of the peak, and θ is the angle. The crystallite sizes of p-$TiO_2$, Cu, Zn-$TiO_2$, and Cu, Zn-$TiO_2$-N are 2.549 nm, 2.505 nm, and 2.501 nm, respectively, showing no significant change. Since there is no peak for the Cu, Zn in the XRD data indicates that Cu, Zn substitutionally doped in $TiO_2$ [15].

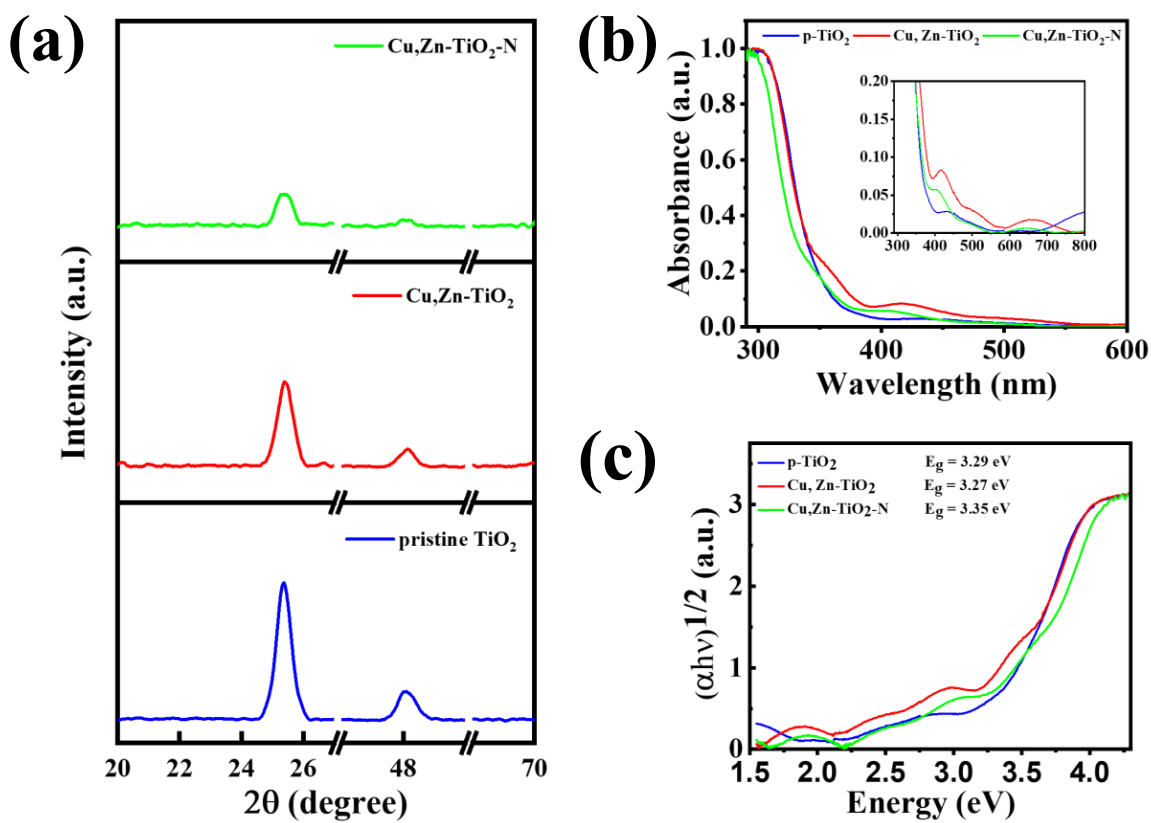

**Fig. 2 (a)** XRD pattern, **(b)** Absorbance Spectra, **(c)** Tauc Plot of pristine and co-doped TiO₂ thin film electrodes.

## Field emission scanning electron microscopy

We have taken the FE-SEM images to probe the morphology of undoped and co-doped $TiO_2$ electrodes. We observed the uniform and flat surface morphology of the pristine $TiO_2$ (Fig. 3(a)). Cu, Zn co-doped $TiO_2$, and Cu, Zn, N doped $TiO_2$ FE-SEM micrograph reveals the significant changes in the morphology (Fig. 3(a), 3(b)). Doping of Cu, Zn, and N creates a porous structure on the surface of the electrode, resulting in increased effective surface area of contact with the electrolyte. The electrodes with porous structure morphology can access a more significant number of interfacial and electrocatalytic active sites. It will expedite the performance of the electrodes. In Fig. 4, the elemental mapping shows the distribution of elements is uniform throughout the sample area.

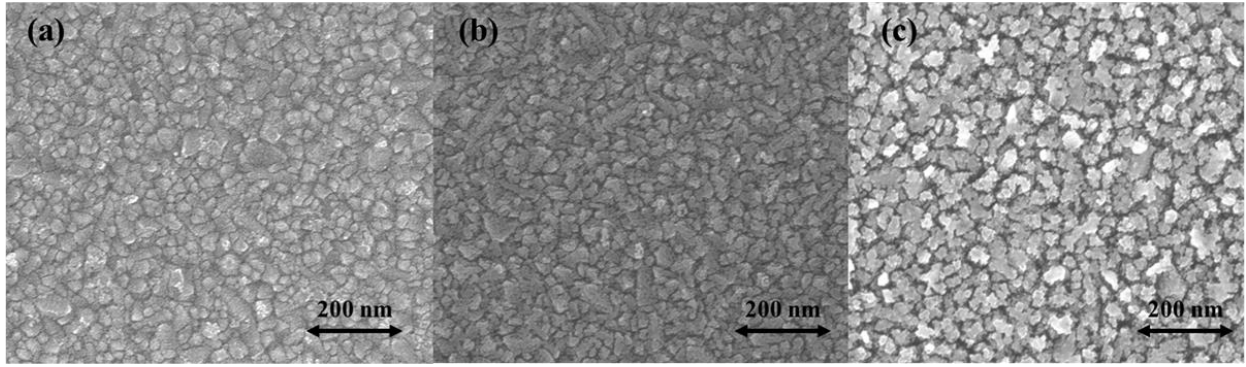

**Fig. 3** FESEM Micrographs of **(a)** pristine TiO$_2$ **(b)** Cu, Zn doped TiO$_2$ **(c)** Cu, Zn, N doped TiO$_2$ electrodes.

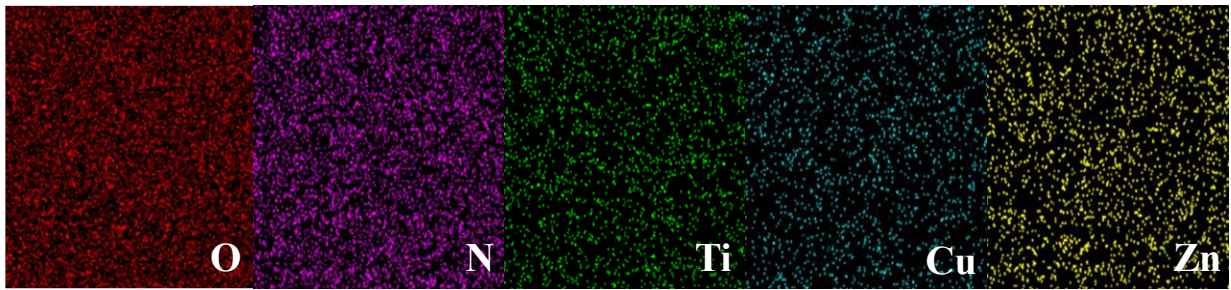

**Fig. 4** Elemental color maps of all the elements present in the synthesized co-doped TiO$_2$ electrodes.

## UV-Visible absorption spectroscopy

Absorption spectrum analysis is the direct way to find out the optical properties, especially the bandgap of the materials. When a photon of specific energy is absorbed by an electron present in a lower energy state, it gets excited and jumps to a higher energy state. In this process, changes in transmitted radiation provide us an idea about the different electronic transitions. To calculate the bandgap of semiconductor materials, Tauc developed a method in 1966 which was further improved by Davis and Mott. His approach is based on the energy-dependent photon absorption coefficient, presented by the equation,

$$(\alpha \cdot h\nu)^{\frac{1}{\gamma}} = B(h\nu - E_g) \quad ...(2)$$

Where $\alpha, h, \nu, E_g$ are the absorption coefficient, plank's constant, absorbed photon's frequency and bandgap energy of the material, respectively. B is the equation constant, and the value of the factor $\gamma$ depends on the nature of the electronic transition. The value of $\gamma$ can be 2 for the indirect band transition and 0.5 for the direct band transition [16]. The absorption of photons below the bandgap energy $E_g$ shows the presence of defects and trap states, usually found in the doped and surface-modified materials. These modifications create intra-band gap states manifested as Urbach tails in the spectrum [16, 17].

UV–Visible absorption spectra were collected (from 200 nm to 800 nm) to check the optical bandgap of the synthesized $TiO_2$ based electrodes (Fig. 2(b), (c)). Absorption spectra (Fig. 2(b)) of co-doped $TiO_2$ thin films show some significant absorption in the visible region due to defects and trap states formation caused by doping. The optical bandgap of the fabricated thin films was calculated using the tauc plot method (Fig. 2(c)). A small change in the optical bandgap was observed in the samples. The bandgap energy of the pure $TiO_2$ is 3.29 eV. The value of bandgap energy for Cu, Zn-$TiO_2$ and Cu, Zn-$TiO_2$-N are 3.27 eV and 3.35 eV, respectively. The increment in the bandgap of these samples is due to Zn doping [18], while the interstitial doping of N causes a negligible effect on bandgap, also confirmed by another group [19].

## Electrochemical Impedance Spectroscopy

Electrochemical-Impedance Spectroscopy is a very advanced technique that uses an AC signal of small amplitude to investigate the impedance-based behavior of the cell. It gives us information on charge transfer dynamics, capacitive and diffusion processes occurring in the electrochemical cell [20-22]. The schematic of carrier kinetics is shown in Fig. (S1). The doping of the Cu, Zn, and N changed the morphology of the surface of the electrodes. The porous structure increased the active surface area for the chemical reaction. The impedance data has been collected from $10^{-1}$ Hz to $10^6$ Hz. In the pristine and co-doped $TiO_2$ electrodes, similar trend is observed in the bode plots (Fig. S2-S4). The prominent increment in the resistance in low frequency region shows the non-ideal capacitive behavior. In the Nyquist plot, two semicircles appear; the first semicircle gives the information on the electron kinetics and recombination time corresponding to high frequency response (Fig. 5). The second semicircle shares information on ion migration and ion recombination time on the electrode/electrolyte interface against the low frequency response. The low frequency arch shows some diffusive nature which supports the migration of ions to the active surface area of the electrodes. In the Nyquist plots (Fig. 5 (a), (b) & (c)), the high-frequency semicircle gets smaller with the co-doping, and the low-frequency semicircle gets bigger. The Nyquist plots contain the semicircles depressed in nature, revealing the deviation from the ideal Debye type charge carrier relaxation process. To discuss the physical process taking place in the pristine and co-doped $TiO_2$, we fitted the impedance data with the equivalent circuit model as shown in Fig. 6. The circuit resistances R1, R2, R3, R4 represent the charge transfer resistance ($R_{ct}$) of the counter electrode, defect states present inside the electrodes, interface, and diffusion, respectively. The circuit capacitances C1, C2, and C3 represent the capacitance of the Pt electrode, the capacitance of the defect states ($C_{ds}$), and double layer capacitance ($C_{dl}$) at the interface of the electrode/electrolyte, respectively. Impedance spectrum consists of three regions: high frequency, intermediate frequency, and low frequency. W is the Warburg coefficient present in the system. The high frequency region conforms to the charge transport and redox reaction at the Platinum electrode (counter electrode). In the intermediate frequency region, electron transport occurs in the electrodes and back reaction to the electrode/electrolyte interface. Doping of Cu, Zn, and N created the defects in $TiO_2$ electrodes, which work as charge recombination centers (trap states), confirmed by the increased capacitance ($C_{ds}$) value in the intermediate frequency region (Table (S1)). The low frequency region is attributed to ion transport in the electroactive layer along with the redox reaction and ion interaction at the electrode/electrolyte interface.

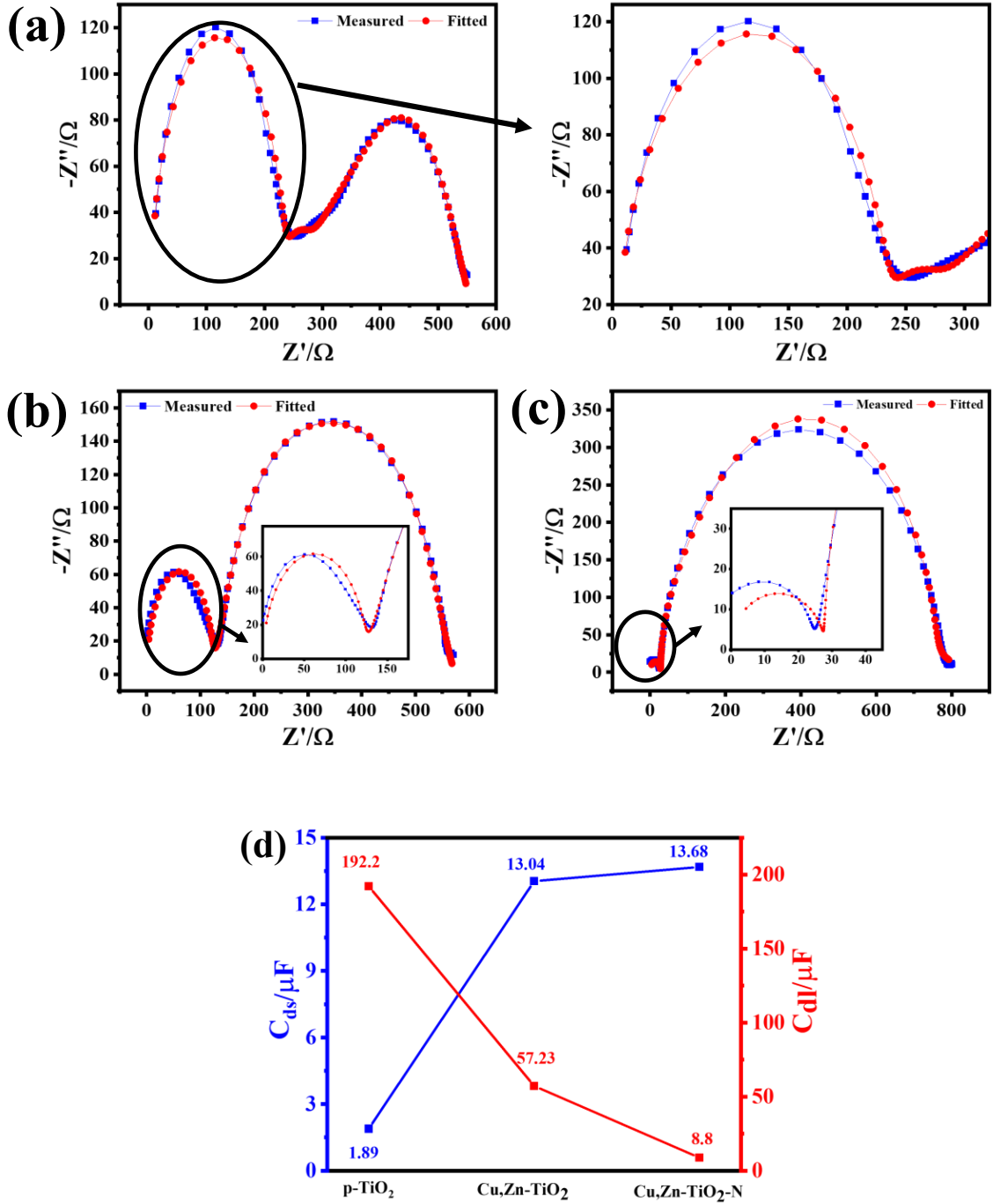

**Fig. 5** Nyquist plots of **(a)** pristine $TiO_2$ **(b)** Cu, Zn co-doped $TiO_2$ **(c)** Cu, Zn, N co-doped $TiO_2$ **(d)** change in the capacitance ($C_{ds}$ and $C_{dl}$) with the doping.

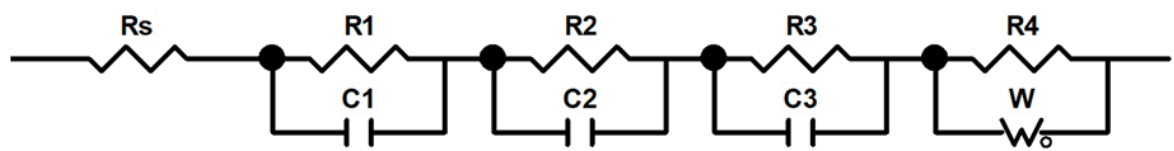

**Fig. 6** Equivalent circuit model.

The diffusion of electroactive carriers through the interface decreases the double layer capacitances at the electrode/electrolyte interface. The polarization resistance in low frequency region is increased with the change in morphology of the electrode with the doping. Increment in the polarization resistance favors the increment in the potential across the electrode/electrolyte interface, which results in a significant decrement in the double layer capacitance of the interface (Table (S1)). The migration of minority charge carriers to the surface states occurs in low frequency response.

## Conclusions

In this research article, we synthesized the pristine and co-doped electrodes using spin coating method. We performed the Thin Film XRD, FESEM, UV-Visible Spectroscopy, and Electro-impedance spectroscopy characterizations of the synthesized samples. Thin film XRD data confirms only the anatase structure of the $TiO_2$ based electrodes. FESEM images manifest the change in surface morphology of the electrodes with the dopants. The change in morphology supports the low contact resistance with the electrolyte and larger surface area of contact. Elemental color mapping detected the presence of the Ti, Cu, Zn, O, and N elements with the uniform distribution. A significant improvement in the visible light absorption is observed with the co-doping. Tauc plot analysis shows the bandgap of p-$TiO_2$, Cu, Zn doped $TiO_2$, and Cu, Zn, and N doped $TiO_2$ are 3.29, 3.27, and 3.38, respectively. Intermediate bandgap states were also observed in the Tauc plot. The EIS measurement is performed to study the charge carrier kinetics in the electrodes. The study concludes that at the defect state capacitance increased due to the localization of the charge carriers and the transport of electrons is enhanced due to the decrement in the charge transfer resistance in bulk. The double layers capacitance decreased, and transportation of the electrons and ions to the surface/interface improved.

## Acknowledgements

MKV wants to acknowledge Mr. Ramesh Kumar, Research Scholar, AREIS Lab, IIT Roorkee for the important discussions regarding the article.

## Declarations

**Conflict of interest:** The authors declare no conflict of interest.

**Supplementary Information:** The online version contains supplementary material available at (Link will be provided by the journal)

Films: Photocatalytic and Bactericidal Activity. ACS Appl. Mater. Interfaces **13**:10480-10489. https://doi.org/10.1021/acsami.1c00304
20. A.R.C. Bredar, A. L. Chown, A. R. Burton, and B. H. Farnum (2020) Electrochemical Impedance Spectroscopy of Metal Oxide Electrodes for Energy Applications. ACS Appl. Energy Mater. **3**:66-98. https://doi.org/10.1021/acsaem.9b01965
21. N. K. Tailor, S. P. Senanayak, M. Abdi-Jalebi, and S. Satapathi (2021) Low-frequency carrier kinetics in triple cation perovskite solar cells probed by impedance and modulus spectroscopy. Electrochimica Acta **386**:138430. https://doi.org/10.1016/j.electacta.2021.138430
22. R. Kumar, J. Kumar, P. Srivastava, D. Moghe, D. Kabra, and M. Bag (2020) Unveiling the Morphology Effect on the Negative Capacitance and Large Ideality Factor in Perovskite Light-Emitting Diodes. ACS Appl. Mater. Interfaces **12**:34265-34273. https://doi.org/10.1021/acsami.0c04489
12